\documentclass[preprint,superscriptaddress]{revtex4}
\usepackage{mathrsfs}
\usepackage{graphicx}
\usepackage{dcolumn}
\usepackage{bm}
\usepackage{amssymb}
\usepackage{amsmath}
\usepackage{paralist}
\usepackage[colorlinks,linkcolor=red,anchorcolor=blue,citecolor=green]{hyperref}

\newcommand{\Tr}{\mathrm{Tr}}
\def\be{\begin{equation}}
	\def\ee{\end{equation}}
\def\bea{\begin{eqnarray}}
	\def\eea{\end{eqnarray}}

\newtheorem{definition}{Definition}

\begin{document}
	\title{Excessive precision compromises accuracy even with unlimited resources due to the trade-off in quantum metrology}
	\author{Cong-Gang Song}
	\affiliation{Innovation Academy for Precision Measurement Science and Technology, Chinese Academy of Sciences, Wuhan 430071, China}
	\affiliation{University of Chinese Academy of Sciences, Beijing 100049, China}
	\author{Qing-yu Cai}
	\thanks{qycai@hainanu.edu.cn}
	\affiliation{Center for Theoretical Physics, School of Physics and Optoelectronic Engineering, Hainan University, Haikou, 570228, China}
	
	\date{\today}

\begin{abstract}

This paper provides a unified definition of precision and accuracy from the perspective of distinguishing neighboring quantum states. We find that the conventional quantum Cram\'er-Rao bound underestimates the effect of statistical noise, because the biases of parameters were inappropriately ignored. Given that probability estimation is unbiased, defining precision based on probability distributions provides a more accurate approach. This leads to a correction of factor 2 to the traditional precision lower bound. The trade-off between precision and accuracy shows that precision can be further improved by sacrificing accuracy, while it should be restricted by inherent precision limit determined by the number of sampling. The inherent precision limit can reach the Heisenberg scaling even without entanglement resources, which, however, comes at the cost of significantly reduced accuracy. We show that increasing sampling may decrease accuracy when one pursues excessive precision, which indicates the trade-off should be considered even with unlimited resources.

\end{abstract}
	
\maketitle


\section*{INTRODUCTION}
	
Accuracy and precision are frequently used to describe the quality of measurement and estimation results, but their meanings often exhibit subtle differences depending on the context. In parameter estimation theory, precision refers to the quality of the estimator, typically quantified by the uncertainty or standard deviation of the parameter. In contrast, accuracy denotes the deviation between the estimator and the true value, usually represented by bias~\cite{RN813,RN271}. It is widely accepted that maximum likelihood estimation can achieve asymptotic unbiasedness~\cite{RN937}, meaning the bias approaches zero. Consequently, precision is often considered sufficient to fully characterize the measurement result, with its lower bound defined by the quantum Cram\'er-Rao bound (QCRB)~\cite{RN271,RN811,RN282}, from which definitions of the standard quantum limit~\cite{RN282,RN308,RN272,RN447,RN295}, the Heisenberg limit~\cite{RN282,RN447,RN295,RN279}, and the super-Heisenberg limit~\cite{RN293,RN292,RN277,RN280,RN469} are derived. In certain practical sensing devices, precision is defined as the sensor's inherent resolution or minimum scale. For example, in atomic clocks, precision is determined by the time interval between two ``ticks", which is fixed by the transition frequency of the system's energy levels. Accuracy can also be defined as the ratio of precision to uncertainty~\cite{RN644}, essentially equivalent to the Allan variance~\cite{RN643,RN648}. At first glance, these definitions may appear quite different, with uncertainty playing distinct roles: in the former, it primarily defines precision, while in the latter, it quantifies accuracy. A key objective of this paper is to revisit the definitions of precision and accuracy and to shed light on the deep connection between these two concepts.

To unify these two definitions within a single framework, this paper addresses the issue from the perspective of probability distributions rather than parameter estimation. We define accuracy $\alpha$ as the degree to which neighboring quantum states can be distinguished, and precision $\delta \varphi$ as the minimum detectable signal that satisfies the quantum state distinguishability condition at a specified level of accuracy. The previously discussed definitions of precision and accuracy are specific manifestations of this unified framework. Moreover, we identify a fundamental trade-off between precision and accuracy, governed by the intrinsic probabilistic and stochastic nature of quantum mechanics. 
This trade-off provides greater flexibility in the design of sensor scale (i.e., precision) for practical devices. For example, one can sacrifice some accuracy to achieve better precision, and vice versa. However, we also note that this approach does not always apply universally. In parameter estimation theory, there exists an irreducible inherent precision, determined by the number of sampling and the initial parameter. In some extreme cases, this inherent precision can reach the Heisenberg limit even without entanglement resources, but as a price, the accuracy even decreases as the number of sampling increases.
	
Defining precision and accuracy directly from the perspective of probability distributions offers two key advantages. First, unbiased parameter estimation is challenging under limited resources, as unbiasedness typically requires specialized unbiased estimators or asymptotically holds only for maximum likelihood estimation in the large $n$. This bias prevents uncertainty $\Delta \varphi$ from fully describing the quality of the measurement. In contrast, estimation of the probability distribution itself is always unbiased, and thus defining precision in terms of the distinguishability of probability distributions helps address this issue. We find that, for the same expected accuracy, a correction factor of 2 must be applied to the traditional precision lower bound.
Second, quantum metrology can improve measurement outcomes by introducing additional measurement resources~\cite{RN153,RN249,RN248,RN267}. From the perspective of parameter estimation, the role of all measurement resources appears to be same and trivial, aimed at reducing noise to minimize statistical errors~\cite{RN282,RN272}. However, from the perspective of probability distributions, the mechanisms by which different resources enhance measurement outcomes are distinct. Only repeated sampling schemes effectively reduce statistical noise, while resources such as entanglement, multi-body product states, nonlinear effects primarily serve to enhance the signal, rather than reduce the noise. These insights provide a clearer physical picture of quantum sensing.

\section*{RESULTS}
\subsection*{Precision and accuracy}
\label{TOBPAA}

From a physical intuition standpoint, the most straightforward definition of precision is the smallest detectable signal that a system can measure. For sensors with built-in scales, signals smaller than this scale cannot be measured, making the scale a direct indicator of precision. For sensors without built-in scales, there is more flexibility in defining precision. It is generally accepted that the minimum detectable signal should at least meet a unit signal-to-noise ratio (SNR)~\cite{RN71}, otherwise, the signal becomes indistinguishable from the noise. Statistical noise can be quantified by the mean squared error:
   \begin{equation}
    	\begin{aligned}
		\mathrm{MSE}(\hat{\varphi})&=E[(\hat{\varphi} -\varphi)^2]
		\\&=E[(\hat{\varphi}-E[\hat{\varphi}] )^2]+(E[\hat{\varphi} ]-\varphi)^2
		\\&=\mathrm{Var(\hat{\varphi})} +\mathrm{Bias} (\hat{\varphi} )^2.
    	\end{aligned}
    \end{equation}
When the sample size $n$ is sufficiently large, maximum likelihood estimation can asymptotically achieve an unbiased parameter estimate, i.e., $\mathrm{Bias} [\hat{\varphi} ]=0$~\cite{RN937}. Therefore, in parameter estimation theory, the uncertainty  $\Delta \varphi =\sqrt {\mathrm {Var}[\hat{\varphi}]}$ is commonly used to describe measurement precision. However, when resources are limited, it is not always possible to ensure that $\mathrm{Bias} [\hat{\varphi} ]=0$, leading to potential inaccuracy in parameter estimation. Considering that in a binomial distribution, frequency-based probability estimation is unbiased, i.e., $E(\hat{p})=E(k/n)=p$, we propose the following definitions from the perspective of probability distributions.
	
	\begin{definition} 
		For quantum sensors with built-in accuracy $\alpha$, precision $\delta \varphi$ is defined as
		\begin{equation}
			\delta \varphi : = \left \{  \delta \varphi=\mathrm{min}|\varphi -\varphi _0| , |p(\varphi)-p(\varphi _0)|\ge \alpha (\Delta p(\varphi)+\Delta p(\varphi _0) ),\alpha \in (0,\infty ) \right \}. 
		\end{equation}
		For quantum sensors with built-in precision  $\delta \varphi$, accuracy $\alpha$ is defined as
		\begin{equation}
			\alpha :=\left \{ \alpha =\frac{|p(\varphi)-p(\varphi _0)|}{\Delta p(\varphi)+\Delta p(\varphi _0) },\delta \varphi =|\varphi -\varphi _0|  \right \}. 
		\end{equation}
	
		Here, $\varphi _0$ represents the initial known parameter, and $\varphi$ is the unknown parameter to be measured, which are respectively encoded in the initial and final quantum states. For any type of quantum state, one can always perform a binary (yes-no) measurement, where the measurement operators consist of only two elements $ \{  E, I- E  \} $. Therefore, $p(\varphi_0 )=\Tr [E \rho_{\varphi_0} ] $, $p(\varphi )=\Tr [E\rho_{\varphi} ] $ represent the measurement probabilities for the initial and final states, respectively, under the same element $E$. They can also represent measurement outcomes projected onto the element $I-E$, but this makes no essential difference in the context of binary measurement.
	\end{definition}
	
In the subsequent sections, we will elucidate the rationale behind this definition by examining it from the perspectives of probability distributions and the distinguishability of neighboring quantum states.
	
For a weak signal, the quantum states to be distinguished are the initial state $|\psi\rangle$ and the final state $|\psi'\rangle$ after interaction with the signal. We can select a suitable basis $\{|0\rangle, |1\rangle\}$ such that $|\psi\rangle = (|0\rangle + |1\rangle)/\sqrt{2}$, and the final state becomes $|\psi'\rangle = (|0\rangle + e^{i \varphi} |1\rangle)/\sqrt{2}$. Using the known initial state as the measurement basis, these quantum states can be maximally distinguished \cite{RN301}, some alternative proofs can also be found in Methods section \ref{TOC}.
The measurement outcomes are described by the equations:
	\begin{equation}\label{eq1}
		\begin{aligned}
			p(|\psi\rangle) &= |\langle \psi | \psi \rangle|^2 = F(|\psi\rangle, |\psi\rangle) = 1, \\
			p(|\psi'\rangle) &= |\langle \psi | \psi' \rangle|^2 = F(|\psi\rangle, |\psi'\rangle) = \frac{1}{2}(1 + \cos \varphi).
		\end{aligned}
	\end{equation}
Here, $F$ represents the relative fidelity between the initial and final states, equating to the probability that one quantum state is measured based on the other. This process can be achieved through a Ramsey measurement \cite{RN282,RN71}. According to Wotters' criteria, two quantum states are considered distinguishable if the difference between their measurement results satisfies \cite{RN301}:
	\begin{equation} \label{SNR}
		|p - p_0| \geq \Delta p + \Delta p_0.
	\end{equation}
This condition implies that the distinguishability of the quantum states depends on the measurement signal exceeding or matching the noise (i.e., $SNR \geq 1$). For projection measurements, which conform to the binomial distribution, $\Delta p = \sqrt{p(1-p)/n}$ for $n$ independent samples.
Substituting Eq. (\ref{eq1}) into Eq. (\ref{SNR})  yields
	\begin{equation} \label{FFODC}
		F \leq \frac{n}{n+1},
	\end{equation}
defining the minimum distance or maximum relative fidelity between two distinguishable quantum states for a given finite resource $n$.
Considering $F = (1 + \cos \varphi)/2$, the precision $\delta \varphi$ can be defined as the minimum detectable signal that satisfies Eq. (\ref{FFODC}), that is
	\begin{equation}\label{MDS}
		\delta \varphi \geq \varphi_{min} = \arccos\left(\frac{n-1}{n+1}\right).
	\end{equation}
For large $n$, $\varphi_{min}$ approximates to $2/\sqrt{n}$, i.e., $\delta \varphi \geq 2/\sqrt{n} $.
The same concept can be directly applied to the parameter $\varphi$, that is $\delta \varphi=\left|\varphi-\varphi_0\right| \geq \Delta \varphi+\Delta \varphi_0$. It is generally assumed that the initial parameter $\varphi_0$ is completely known, so we have  $\Delta \varphi_0 =0$. This leads to the traditional QCRB $\delta \varphi \geq 1/\sqrt{nF_Q(\varphi)}=1/\sqrt{n}$. Notably, this differs from Eq. (\ref{MDS}) by a factor of 2.
We will explain the reason for this correction factor at the end of this section.

\begin{figure*}[!hbt]
		\centering
		\includegraphics[width=16cm]{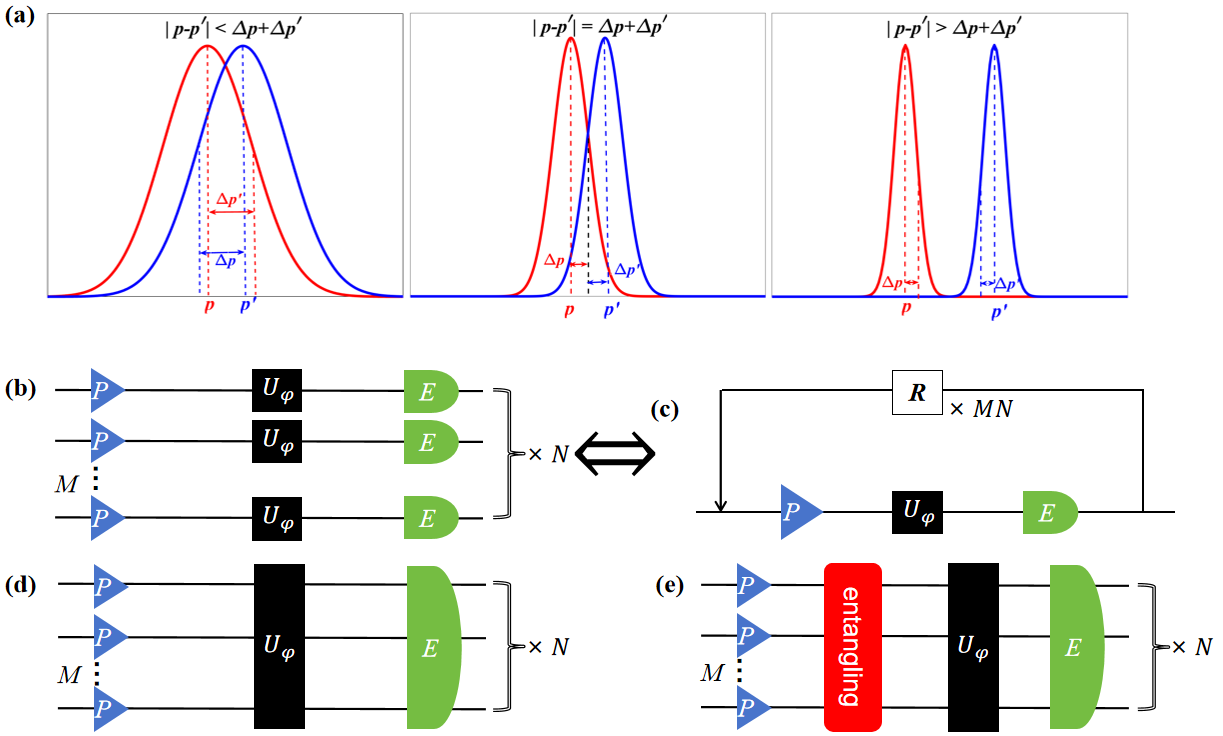}
		\caption{ Basic strategies and principles of quantum metrology. (a) The capability to distinguish neighboring quantum states varies with different SNRs. (b) Multiple quantum states independently probe the signal. (c) Repeated measurements enhance signal detection. Note that panels (b) and (c) serve similar purposes in parameter estimation. (d) Multiple quantum states are utilized as a product state to probe the signal. (e) A many-body entangled state is employed as the probe state. Notation: the ``P" box represents the preparation of the initial state, the ``E" box denotes the measurement and estimation of the final state, and the ``R" box indicates the reset of the final state.}\label{SNR3}
	\end{figure*}

Indeed, $SNR \geq 1$ is not an absolute criterion for distinguishing neighboring quantum states. As shown in Fig.~\ref{SNR3}(a), quantum states can still be partially discerned even when the SNR is below 1. While using this threshold to differentiate neighboring quantum states can enhance measurement precision, it may adversely affect distinguishability, thus compromising the credibility and accuracy of the results. Conversely, setting an SNR greater than 1 as the criterion improves accuracy but can reduce precision. This highlights the challenge of simultaneously improving precision and accuracy due to limited resources, driven by the fundamental inability to perfectly distinguish between two non-orthogonal quantum states. This challenge is rooted in the probabilistic and stochastic aspects of standard quantum mechanics, stemming from the superposition principle and the measurement postulate.

To quantitatively describe the precision and accuracy within a framework, we introduce a parameter $\alpha$ and modify Eq.(\ref{SNR}) to account for this variability:
	\begin{equation} \label{SNRa}
		|p - p_0| \geq \alpha(\Delta p + \Delta p_0).
	\end{equation}
Additionally, Eqs.(\ref{FFODC}) and (\ref{MDS}) are adapted to incorporate $\alpha$ as follows:
	\begin{equation} \label{FSNRa}
		F \leq \frac{n}{n + \alpha^2},
	\end{equation}
	\begin{equation}\label{precision}
		\delta \varphi \geq \arccos\left(\frac{n - \alpha^2}{n + \alpha^2}\right).
	\end{equation}
Here, the range of $\alpha \in (0, \infty)$. When $\alpha \rightarrow \infty$, only two orthogonal quantum states can satisfy the distinguishability condition in Eq.(\ref{SNRa}), setting the measurement precision to the lowest level (only the signal that drives the initial state to evolve into the orthogonal final state can be detected and recorded) while achieving the best accuracy (i.e., the two quantum states can be reliably distinguished). On the contrary, when $\alpha \rightarrow 0$, any two quantum states can satisfy the distinguishability condition. In this case, the precision can be very high, but the accuracy is greatly reduced, because even two identical quantum states can satisfy the condition, which is obviously impractical. When $\alpha = 1$, the original Eq.(\ref{SNR}) condition is restored.

Eq.(\ref{FSNRa}) can also be reformulated as:
	\begin{equation} 
		1 - F \geq \frac{\alpha^2}{n + \alpha^2}.
	\end{equation}
Here, the left side represents the $signal = \delta p = 1 - F$, disregarding the higher-order terms of $O(\delta \varphi^3)$, it can be deduced that $\delta p = 1 - |\langle \psi | \psi' \rangle|^2 \approx (\delta \varphi \triangle G)^2$, which is also expressed by the Fubini-Study metric $\delta p = ds^2/4 = F_Q(\varphi) \cdot \delta \varphi^2/4$ \cite{RN327,RN288}. Here, $G$ acts as the signal generator, such that $|\psi' \rangle = e^{i \varphi G} |\psi \rangle$. From this, we derive:
	\begin{equation}\label{trade1}
		\delta \varphi \geq \frac{2\alpha}{\sqrt{n + \alpha^2} \sqrt{F_Q(\varphi)}}.
	\end{equation}
	When $\alpha^2 \ll n$, Eq.(\ref{trade1}) simplifies to:
	\begin{equation}\label{trade2}
		\delta \varphi \cdot \frac{1}{\alpha} \geq \frac{2}{\sqrt{n F_Q(\varphi)}},
	\end{equation}
indicating a trade-off between precision and accuracy. Notably, the right-hand side is proportional to $\sqrt{n F_Q(\varphi)}$, aligning with the conventional QCRB. 
Neglecting higher-order terms, we can explicitly define precision and accuracy directly from the perspective of parameters, such that 
	\begin{equation}
		\delta \varphi : = \left \{  \delta \varphi \ge 2\alpha/\sqrt{nF_Q(\varphi )},\alpha \in (0,\infty ) \right \}, 
	\end{equation}
	\begin{equation}
		\alpha :=\left \{ \alpha = \delta \varphi \sqrt{nF_Q(\varphi )}/2 ,\delta \varphi =|\varphi -\varphi _0|  \right \}.
	\end{equation}
From this definition, the role of quantum Fisher information in enhancing both precision and accuracy becomes more intuitively evident.
	
Similar to the case in Eq. (\ref{MDS}), under the same accuracy expectation (e.g., $\alpha=1$), the QCRB as a precision lower bound requires a factor of 2 correction. This is because the parameter and the probability $p$ are not simply linearly related, and the commonly used error propagation formula $\Delta \varphi=\Delta p \cdot| d\varphi/dp|$ does not strictly apply in this context. Thus the unbiased estimate of $p$ does not naturally extend to an unbiased estimate of the parameter $\varphi$. In other words, because of the estimation bias of the parameter, using the QCRB as the minimum detectable signal actually underestimates the magnitude of the statistical noise, which is the origin of the factor of 2 correction. It is easy to verify that if the parameter $\varphi$ and probability $p$ satisfy a linear relationship, both $\varphi$ and $p$ become unbiased estimators, and the factor of 2 correction naturally vanishes. In fact, all parameter estimation is based on probability distributions, so defining precision directly from the probability distribution not only resolves the ``bias" issue but is also conceptually more fundamental.
	
On the other hand, we can view traditional parameter estimation theory as equivalent to a quantum sensor with an inherent accuracy of $\alpha=0.5$ (which is, of course, lower than our ideal expectation). In this case, the expression for the precision lower bound returns to the traditional QCRB. Of course, in practical scale design, we can flexibly adjust the allocation between precision and accuracy. For instance, we could design a precision scale larger than the QCRB, thereby improving the accuracy $\alpha$ to meet our ideal expectation.
For sensors with built-in precision, such as atomic clocks,  the precision is fixed and represents the time interval between two ticks. The definition of accuracy $\alpha$ here is entirely consistent with that in the Ref.\cite{RN644,RN643}. In this case, increasing the quantum Fisher information or reducing the uncertainty does not affect the precision, but improves the accuracy. However, using a finer resolution (higher precision) with the same resources inevitably sacrifices accuracy. Similar trade-offs between precision and accuracy in time measurements have been identified in previous work~\cite{RN807}, attributed to the randomness of thermal processes. In contrast, the trade-off we identify here arises from quantum projection noise, stemming from the intrinsic randomness of quantum mechanics. Ultimately, the definitions of precision and accuracy can coexist seamlessly within our framework. The difference lies in one approach embeds precision, while the other embeds accuracy. Given limited resources, enhancing precision will inevitably reduce accuracy, and vice versa.

The above discussion can also be extended to POVM measurements $ \{  E_0, E_1, \dots E_{N-1} \} $. However, since the measurement outcomes now form a multinomial distribution, the quantum state distinguishability condition Eq.(\ref{SNRa}) cannot be directly applied.  An alternative quantum state distinguishability condition  is as follows \cite{RN301}
		\begin{equation}\label{MSNR}
			 \sqrt{n} \cdot \sqrt{\sum_{i=0}^{N-1} \frac{\left(p_i^{\prime}-p_i\right)^2}{p_i^{\prime}}}\ge \alpha.
		\end{equation}
 where $p_i=\Tr [E_i \rho_{\varphi_0} ]$, $p^\prime _i=\Tr [E_i \rho_{\varphi} ]$ denote the outcome distributions obtained from POVM measurements on the initial and final quantum states, respectively, and $p^\prime _i$ encodes information about the parameter to be estimated. The key challenge is then to identify the optimal POVM measurement. A reasonable conjecture is that the optimal POVM should include the projection operator $E_0=\left|\varphi_0\right\rangle\left\langle\varphi_0\right|$ of the initial state $\left|\varphi_0\right\rangle$, while the remaining POVM elements are orthogonal to $E_0$. Such a POVM operator can completely eliminate the noise and uncertainty associated with the initial state measurement. The remaining task is then to identify the optimal POVM for measuring the final state. This POVM can be expressed as 
		\begin{equation}\label{MBV}
			\left\{E_0=\left|\varphi_0\right\rangle\left\langle\varphi_0\right|, E_{\perp}^1, E_{\perp}^2 \cdots E_{\perp}^{N-1}\right\}, E_{\perp}^i=\left|\varphi_0\right\rangle_{\perp}^i\left\langle\varphi_0\right|.
		\end{equation}
For the initial state, it is clear that $p_0=\Tr[ E_0\rho _0]=1$, $p_{i\ne 0}=\Tr[E_{\perp}^i\rho _0]=0$, and for the final state $p^\prime _0=\Tr[E_0\rho _\varphi ]=F$, $p^\prime _{i\ne 0}=\Tr[E_{\perp}^i\rho _\varphi ]=q_i $.
 Substituting these results into Eq.(\ref{MSNR}), we obtain
 \begin{equation}
 	\sqrt{n}\cdot\left [ \frac{(1-F)^2}{F} +\sum_{i=1}^{N-1} q_i \right ]^{\frac{1}{2} } \ge \alpha .
 \end{equation}
Considering  $\sum_{i=1}^{N-1} q_i=1-F$, the formula can be simplified to $F\le n/(n+\alpha ^2)$, which is our main conclusion Eq.(\ref{FSNRa}). The corresponding trade-off relationship Eq.(\ref{trade2}) also naturally holds. Therefore, under a specific accuracy $\alpha$, the obtained measurement precision $\delta \varphi$ is also consistent with QCRB, that is, for a weak signal, Eq.(\ref{MBV}) is a set of optimal measurement basis sets.
Similar to the binomial case, the probability of the multinomial distribution is also an unbiased estimator, but the parameter estimator itself is biased. Therefore, the factor of 2 correction is still required.

\subsection*{The inherent precision in parameter estimation theory} 
\label{inherent precision}
	
In Methods section \ref{TOC}, we demonstrate that the optimal measurement strategy is not unique. Beyond the method discussed in previous section, which uses the initial state as the measurement basis, there are alternative measurement bases described in Eq.~(\ref{OPMB}) that can also maximally distinguish two neighboring quantum states. Under these bases, the measurement precision is given by  $\delta \varphi \ge 2\alpha /\sqrt{n} $. This is consistent with the conclusions in previous section,  and thus exhibits the same trade-off relationship between precision and accuracy.

The trade-off relationship suggests that, in theory, precision can be infinitely improved at the expense of accuracy. However, in this section, we clarify that, much like actual sensors that have built-in scales or minimum precision, parameter estimation theory also implicitly contains an inherent minimum precision. In other words, with limited resources, regardless of how much accuracy is sacrificed, there will still be a minimum precision.
	
\begin{figure}[!hbt]
		\centering
		\includegraphics[width=14cm]{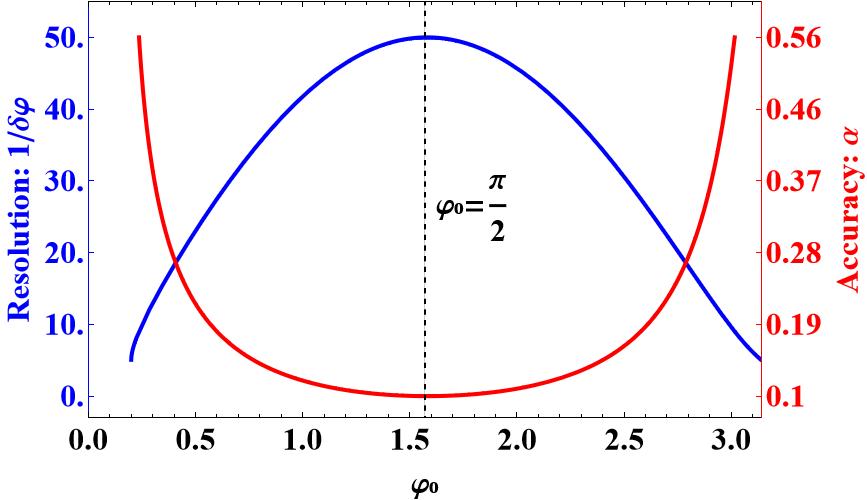}
		\caption{The resolution (i.e. the inverse of precision $1/\delta \varphi$) and accuracy $\alpha$ vary with the initial parameter $\varphi _0$. When $\varphi _0=\pi/2$, the inherent precision is optimal, but the accuracy is the worst. Here, the sample size is set to $n=100$.}\label{PATO}
\end{figure}

This inherent precision limit arises from the constraint on the number of sampling $n$. The measurement and estimation of parameters depend on the readout of probabilities, which are typically estimated from frequency and sample size $n$. Thus, with a finite number of sampling, there exists a minimum detectable probability signal, denoted by $\delta p=1/n$. If the signal is too weak, it fails to induce a significant change in the probability distribution, making it impossible to extract meaningful information from the measurement. For the additional optimal measurement bases proposed in Methods section \ref{TOC}, which correspond to setting the initial parameter $\varphi _0$ to a non-zero value, the minimum signal can be expressed as  $|p(\varphi _0+\delta \varphi )-p(\varphi _0)|= 1/n$, leading to the expression for the inherent precision:
	\begin{equation}
		\delta \varphi =\varphi _0-\arccos\left (\frac{2}{n}+\cos \left(\varphi _0\right) \right),
	\end{equation}
    where $\varphi+\varphi_0 \in(0,\pi)$. Clearly, this inherent precision depends not only on the measurement resources $n$ but also on the choice of the initial value $\varphi_0 $. The corresponding accuracy is given by
    \begin{equation}
    	\alpha =\frac{\delta \varphi \sqrt{n} }{2}. 
    \end{equation}
As shown in Fig.~\ref{PATO}, as the inherent precision enhances, the accuracy decreases, further illustrating the trade-off between precision and accuracy. It is worth noting that when  $\varphi _0=\pi/2$, the inherent precision reaches its optimal value: $\delta \varphi_\mathrm{min}  =\pi/2-\arccos (2/n)\simeq 2/n $. For a given number of sampling $n$, the precision limit exhibits a scaling relationship similar to the Heisenberg limit even without entanglement. However, obtaining this Heisenberg-like precision comes at a cost, as the accuracy is minimized, $\alpha=1/\sqrt{n}$. Notably, as resources increase, the inherent precision limit improves in a manner exceeding our expectations, but the accuracy actually decreases as resources are increased. This suggests that in certain extreme cases, excessively pursuing precision may not be the optimal strategy. Balancing both precision and accuracy is a more effective approach to evaluating parameter information.

\subsection*{The role of measurement resources}\label{TROMR}
	
Quantum metrology offers a significant advantage in that the precision can be substantially improved by augmenting the resources utilized. These resources include the number of repeated sampling, ensembles of multiple quantum states, the exotic quantum states such as entangled or squeezed states, as well as nonlinear effects~\cite{RN282}. Theoretically, the potential for enhancing precision is limitless. With sufficient increases in measurement resources, precision can be improved to approach infinitesimally small values.
	
Conventionally, it is recognized that the enhancement in precision primarily stems from the ability of increased measurement resources to suppress noise, thus reducing statistical errors~\cite{RN272,RN282}. However, when viewed from the perspective of probability distributions, a different insight can be gained. Reconsidering $ \text{Eq.}(\ref{SNR}) $, where the left side, $ \delta p = 1 - F = \frac{F_Q(\varphi) \cdot \delta \varphi^2}{4} $, represents the signal, and the right side, $ \Delta p = \sqrt{\frac{F(1-F)}{n}} $, represents the noise. Moreover, for a weak signal $ \varphi $, it is typically the case that $ F \in (0.5, 1) $, within which $ \Delta p $ increases as $ F $ decreases. In this range, increasing $ F_Q(\varphi) $ or decreasing $ F $ not only fails to suppress noise but may actually amplify it. The improvement in precision under such conditions is due to signal enhancement rather than noise reduction.  Therefore, it is crucial to explore the role of different measurement resources in different scenarios.
	
Performing $N$ repeated measurements inevitably consumes extra time, so it is classified as a temporal resource. Increasing the number of quantum states $M$ requires the use of more particles. The increase in the number of particles leads to a larger volume and density, so it is considered a spatial resource. In addition, quantum entanglement involving complex interactions of quantum states is a unique and extremely valuable quantum resource.
	
The spatial resource $M$ and the temporal resource $N$ have not been strictly distinguished in previous works. This is because measuring independent identical quantum states from an ensemble is equivalent to repeated sampling, as shown in Fig. \ref{SNR3} (b) and (c). The role of both types of resources is essentially the same, thus, $n = MN$, and the corresponding expressions for signal and noise are as follows:
	\begin{equation}	
		\delta p=  \frac{1}{2} \left(1 - \cos \varphi\right), \quad \Delta p=\frac{\sin \varphi}{2\sqrt{MN}}.
	\end{equation}
In this context, spatial and temporal resources contribute equally to reducing noise and statistical errors. The precision is governed by Eq.~(\ref{MDS}).
	
On the other hand, the direct product state of $M$ independent particles can also be treated as a whole in the quantum metrology process, as illustrated in Fig.\ref{SNR3}(d). Here, the prepared initial state can be described as $\left|\Psi_0\right\rangle^{\otimes M} = |+\rangle^{\otimes M}$. When the measurement process is repeated $N$ times, the following can be deduced:
	\begin{equation}\label{F2}
		\delta p = 1 - \cos^{2M} \left(\frac{\varphi}{2}\right), \quad \Delta p = \sqrt{\frac{F(1-F)}{N}} \approx \frac{\sqrt{M} \varphi }{2\sqrt{N} } .
	\end{equation}
The precision is then given by
	\begin{equation}
		\delta \varphi \geq 2 \arccos \left(F_0\right)^{1 / 2M} \approx \frac{2}{\sqrt{MN}}
	\end{equation}
where $F_0 = N/(N + 1)$ denotes the critical fidelity required to meet the distinguishability condition. From these calculations, it can be seen that spatial resources, when applied to weak signals, may actually increase noise. The enhancement in precision, however, arises because $M$ direct product states accumulate more phase information per unit time compared to a single quantum state, thereby enlarging the distinguishability between the initial and final states. Consequently, this accelerated evolution allows weaker signals, corresponding to higher precision, to fulfill the quantum state distinguishability condition within the same measurement scheme.
Similarly, for an entangled state of $M$ particles (as depicted in Fig.\ref{SNR3}(e)), the state can be expressed as:
	\begin{equation}
		\left|\Psi_0\right\rangle^{(M)} = \left(|0\rangle^{\otimes M} + |1\rangle^{\otimes M}\right) / \sqrt{2}.
	\end{equation}
This leads to the following expressions for signal and noise:
	\begin{equation}\label{F3}
		\delta p =  \frac{1}{2} \left(1 - \cos M\varphi\right), \quad \Delta p= \frac{\sin(M\varphi)}{2\sqrt{N}}.
	\end{equation}
The precision is then quantified as:
	\begin{equation}
		\delta \varphi \geq \frac{2}{M} \arccos \left(F_0\right)^{1 / 2} \approx \frac{2}{M\sqrt{N}}
	\end{equation}
Evidently, quantum entanglement does not directly reduce noise. However, compared to the direct product states $ \left|\Psi_0\right\rangle^{\otimes M} $, entangled states $ \left|\Psi_0\right\rangle^{(M)} $ exhibit stronger signal strength. This is because the accelerated evolution of entangled states enhances the distinguishability between the initial and final states, thereby improving the quality of quantum metrology. The phenomenon of entanglement-accelerated evolution has been demonstrated in several studies~\cite{RN304,RN316,RN370}.

Moreover, from the perspective of parameter estimation, quantum Fisher information is intricately linked to the speed of quantum state evolution \cite{RN370,RN291,RN361,RN401,RN402,RN390}, suggesting that increasing quantum Fisher information results in faster evolution. The role of quantum entanglement becomes more intuitively evident in the presence of decoherence. For instance, a study \cite{RN537} shows that using $n$-body maximally entangled states during decoherence can reduce the optimal interrogation time by a factor of $n$. As a result, with a constant total measurement time, the measurement outcome improves by a factor of $1/\sqrt{n}$. However, this benefit is offset by the accelerated rate of decoherence.
	
When dealing with multiple independent identical quantum states, they can be conceptualized as either an ensemble or a product state. Each approach offers distinct advantages for enhancing measurement precision. In the ensemble scenario, spatial resources $M$ and temporal resources $N$ serve equivalent roles in parameter estimation, effectively reducing statistical noise. In contrast, treating the $M$ quantum states as a direct product state increases the speed of evolution, thereby widening the disparity between the initial and final states and enhancing the signal. The introduction of additional quantum resources, such as quantum entanglement, further accelerates this evolution, significantly increasing signal strength and thereby improving precision.
	
Our work also offers new insights into nonlinear quantum metrology schemes. In existing research, two main types of nonlinear metrology schemes are prevalent. The first type introduces nonlinear effects during the beam-splitting and recombination stages, which can be implemented through degenerate parametric amplification, four-wave mixing, and other techniques \cite{RN447,RN951}. This approach is essentially equivalent to preparing an entangled or squeezed state. Therefore, the nonlinear effect here, similar to previous analyses, enhances the signal through specific states. Experimental implementations of this scheme have been achieved on platforms such as optics \cite{RN949,RN950} and Bose-Einstein condensates (BEC) \cite{RN726,RN947}, demonstrating precision beyond the standard quantum limit. However, this type of scheme remains ultimately constrained by the Heisenberg limit.
The second type introduces nonlinear effects during the interrogation stage, where the quantum state interacts with the signal source ~\cite{RN293,RN292,RN277,RN280,RN469}. A typical method involves introducing a nonlinear operator $f(G)$ as the signal generator, i.e.$\left | \psi^\prime  \right \rangle =e^{i\varphi f(G)}\left | \psi  \right \rangle$. This approach can, in principle, surpass the Heisenberg limit, achieving scaling of the form $\delta \varphi \propto n^{-k}, k > 1$. This is achieved by enhancing the signal rather than reducing the noise, and thus without any risk of violating the fundamental laws of quantum mechanics. Current research suggests that this type of metrology scheme can be realized through multi-body interactions \cite{RN532,RN826} or Kerr effects \cite{RN293,RN948}.

\section*{DISCUSSION}
\label{Discussion}

In previous works, the assessment of quantum metrology has primarily focused on the calculation and discussion of quantum Fisher information. However, under finite resources, issues such as whether the QCRB can be saturated, and whether excessive pursuit of precision incurs additional costs, have not received sufficient attention. From the perspective of probability distributions, we point out that the traditional precision lower bound requires a correction factor of 2 in order to meet ideal accuracy expectations. Furthermore, the trade-off between precision and accuracy provides more flexibility in the design of quantum sensors, enabling a dynamic allocation of precision and accuracy according to practical needs. However, we also note that an excessive focus on precision can severely compromise accuracy. In extreme cases, accuracy may even decrease as measurement resources increase.
	
Our method provides a reliable evaluation of both bias and variance, while naturally recovering the conventional QCRB framework. Notably, our method remains rooted in the frequentist perspective and is computationally simple. In contrast, Bayesian methods take a different route by directly optimizing the mean squared error without assuming unbiased estimators. These methods can provide more efficient and accurate parameter estimation under limited data \cite{RN1057,RN1058}. Some Bayesian methods also propose the need to make a constant factor correction to QCRB after excluding the influence of some implicit prior information \cite{RN604,RN281,RN889}, thereby providing a more practical lower bound for parameter estimation. Although Bayesian techniques perform well in limited data and adaptive strategies, they often require additional computational resources. 
An interesting question is whether our framework can be extended to the Bayesian approach. For instance, the accuracy parameter $\alpha$ introduced in this work may serve as a measure of the reliability of prior information.  
Additionally, our analysis naturally yields an upper bound on the fidelity between neighboring quantum states, it may be possible to connect our results with fidelity-based Ziv-Zakai bounds constructed from hypothesis testing \cite{RN922,RN303}. These questions are promising and will be considered in future work.

\section*{METHODS}
\subsection*{The optimal scheme to distinguish two quantum states} \label{TOC}
	
\begin{figure}[!hbt]
		\centering
		\includegraphics[width=8cm]{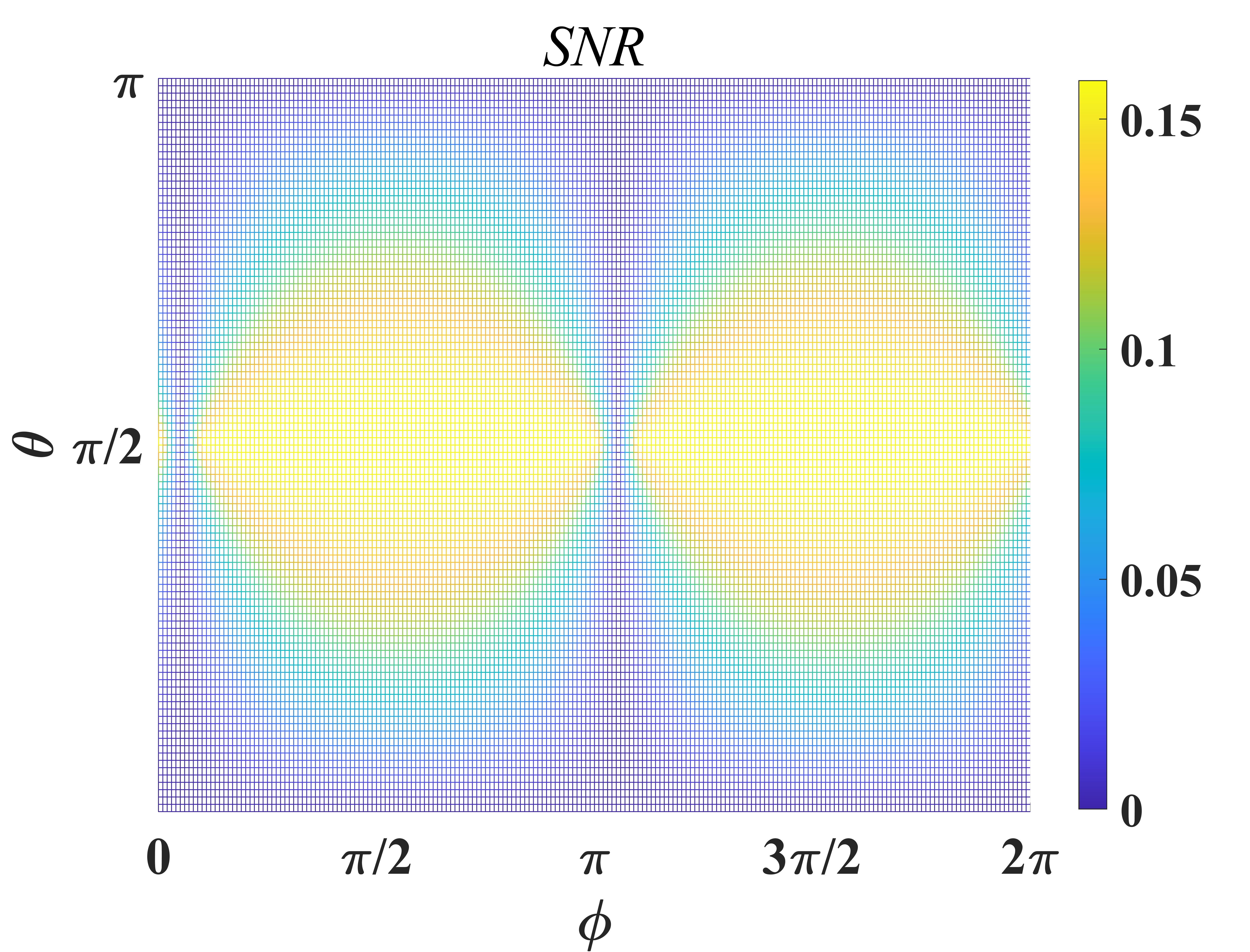}
		\caption{The signal-to-noise ratio for different measurement bases. Here we take $n=1$ and $\varphi=\pi/10$.}\label{SNR mb}
	\end{figure}

The central problem of quantum metrology is to distinguish the known input state $|\psi\rangle $ and the output state $|\psi^{\prime}\rangle=e^{-i \varphi G}|\psi\rangle$ after the interaction with the signal. By choosing an appropriate basis vector, it can always be given
	\begin{equation}
		\begin{aligned}
			|\psi\rangle&=\frac{1}{\sqrt{2}}(|0\rangle+|1\rangle) \\
			\left|\psi^{\prime}\right\rangle&=\frac{1}{\sqrt{2}}\left(|0\rangle+e^{i \varphi}|1\rangle\right).
		\end{aligned}
	\end{equation}
Any measurement basis can be expressed as
	\begin{equation}
		|M\rangle=\cos \frac{\theta}{2}|0\rangle+\sin \frac{\theta}{2} e^{i \phi}|1\rangle.
	\end{equation}
where $|M\rangle$ and $|M\rangle_{\perp} $ constitute a complete measurement basis. Utilizing this set of measurement bases to measure the two quantum states $|\psi\rangle $ and $|\psi^{\prime}\rangle$, the resulting probabilities are:
	\begin{equation}
		\begin{aligned}
			p(|M\rangle,|\psi\rangle) &=|\langle M \mid \psi\rangle|^{2}=\frac{1}{2}(1+\sin \theta \cos \phi) \\
			p\left(|M\rangle,\left|\psi^{\prime}\right\rangle\right) &=\left|\left\langle M \mid \psi^{\prime}\right\rangle\right|^{2}=\frac{1}{2}(1+\sin \theta \cos (\varphi-\phi)).
		\end{aligned}
	\end{equation}
The uncertainties of the measurement results are:
	\begin{equation}
		\begin{aligned}
			\Delta p(|M\rangle,|\psi\rangle) &=\frac{1}{2\sqrt{n}} \sqrt{1-\sin ^{2} \theta \cos ^{2} \phi} \\
			\Delta p\left(|M\rangle,\left|\psi^{\prime}\right\rangle\right) &=\frac{1}{2\sqrt{n}} \sqrt{1-\sin ^{2} \theta \cos ^{2}(\varphi-\phi)}.
		\end{aligned}
	\end{equation}
The signal-to-noise ratio is then given by
	\begin{equation}
		SNR=\left|\frac{\sqrt{n} \cdot \sin \theta \cdot(\cos \phi-\cos (\varphi-\phi))}{\sqrt{1-\sin ^{2} \theta \cos ^{2} \phi}+\sqrt{1-\sin ^{2} \theta \cos ^{2}(\varphi-\phi)}}\right|.
	\end{equation}	
It can be demonstrated that when the measurement bases parameters are 
	\begin{equation} \label{OPMB}
		\left \{ \theta=\frac{\pi}{2},\phi=[\varphi, \pi] \cup[\varphi+\pi, 2 \pi] \right \} .
	\end{equation}
the SNR is maximized, i.e., $SNR_{max}=|\tan (\varphi / 2)|$, as depicted in Fig. \ref{SNR mb}. This parameter set ($\theta,\phi$) is optimal, encompassing the two quantum states themselves and their orthogonal states. Typically, the known quantum state $|\psi\rangle$ is chosen as the measurement basis for convenience.
If $SNR_{max} \geq \alpha$ is used as distinguishability condition, we can get
	\begin{equation}
		\delta \varphi \ge 2\arctan (\frac{\alpha }{\sqrt{n} }  )= \arccos\left(\frac{n - \alpha^2}{n + \alpha^2}\right)
	\end{equation}
this leads to the result consistent with Eq. (\ref{precision}), implying that the precision and accuracy exhibit the same trade-off relationship as in Eq. (\ref{trade2}).
	
From a parameter estimation perspective, the optimal measurement is characterized by the maximum Fisher information. Therefore, a measurement scheme attains optimality if it achieves quantum Fisher information. For instance, the density operator of the output state is $\rho_{\varphi}=\left|\psi^{\prime}\right\rangle\left\langle\psi^{\prime}\right|$. The Symmetric Logarithmic Derivative (SLD) for the pure state is $L=2 \partial_{\varphi} \rho_{\varphi}$ \cite{RN252}, yielding quantum Fisher information $F_{Q}=tr (\rho_{\varphi}L^2)=1$. If the measurement bases described by Eq. (\ref{OPMB}) is used, the classical Fisher information can be calculated as 
	\begin{equation}
		F_c= {\sum_{|\psi\rangle,|\psi\rangle_{\perp}}} \frac{1}{p} \left ( \frac{\partial p}{\partial \varphi}  \right )^2=1 .
	\end{equation}
Evidently, $F_Q = F_C$, which indirectly confirms the optimality of the measurement bases Eq. (\ref{OPMB}) .
	
\section*{DATA AVAILABILITY }
No datasets were generated or analysed during the current study.

\section*{ACKNOWLEDGEMENTS}
This work is supported by NSAF (Grant No.U2130205) and Innovational Fund for Scientific and Technological Personnel of Hainan Province (Grant No.KJRC2023B11).

\section*{AUTHOR CONTRIBUTIONS}
All authors made a significant contribution to the work and were involved in interpreting the results and writing the manuscript.

\section*{COMPETING INTERESTS}
The authors declare no competing interests.


\begin{thebibliography}{10}
	\expandafter\ifx\csname url\endcsname\relax
	\def\url#1{\texttt{#1}}\fi
	\expandafter\ifx\csname urlprefix\endcsname\relax\def\urlprefix{URL }\fi
	\providecommand{\bibinfo}[2]{#2}
	\providecommand{\eprint}[2][]{\url{#2}}
	
	\bibitem{RN813}
	\bibinfo{author}{Bevington, P.~R.}, \bibinfo{author}{Robinson, D.~K.} \&
	\bibinfo{author}{Bunce, G.}
	\newblock \bibinfo{title}{Data reduction and error analysis for the physical
		sciences, 2nd ed}.
	\newblock \emph{\bibinfo{journal}{Am. J. Phys.}} \textbf{\bibinfo{volume}{61}},
	\bibinfo{pages}{766--767} (\bibinfo{year}{1993}).
	
	\bibitem{RN271}
	\bibinfo{author}{Braunstein, S.~L.} \& \bibinfo{author}{Caves, C.~M.}
	\newblock \bibinfo{title}{Statistical distance and the geometry of quantum
		states}.
	\newblock \emph{\bibinfo{journal}{Phys. Rev. Lett.}}
	\textbf{\bibinfo{volume}{72}}, \bibinfo{pages}{3439--3443}
	(\bibinfo{year}{1994}).
	
	\bibitem{RN937}
	\bibinfo{author}{Fisher, R.~A.}
	\newblock \bibinfo{title}{Theory of statistical estimation}.
	\newblock \emph{\bibinfo{journal}{Math. Proc. Cambridge Philos. Soc.}}
	\textbf{\bibinfo{volume}{22}}, \bibinfo{pages}{700--725}
	(\bibinfo{year}{1925}).
	
	\bibitem{RN811}
	\bibinfo{author}{Helstrom, C.~W.}
	\newblock \bibinfo{title}{Quantum detection and estimation theory}.
	\newblock \emph{\bibinfo{journal}{J. Stat. Phys.}}
	\textbf{\bibinfo{volume}{1}}, \bibinfo{pages}{231--252}
	(\bibinfo{year}{1969}).
	
	\bibitem{RN282}
	\bibinfo{author}{Giovannetti, V.}, \bibinfo{author}{Lloyd, S.} \&
	\bibinfo{author}{Maccone, L.}
	\newblock \bibinfo{title}{Advances in quantum metrology}.
	\newblock \emph{\bibinfo{journal}{Nat. Photonics}}
	\textbf{\bibinfo{volume}{5}}, \bibinfo{pages}{222--229}
	(\bibinfo{year}{2011}).
	
	\bibitem{RN308}
	\bibinfo{author}{Caves, C.~M.}
	\newblock \bibinfo{title}{Quantum-mechanical radiation-pressure fluctuations in
		an interferometer}.
	\newblock \emph{\bibinfo{journal}{Phys. Rev. Lett.}}
	\textbf{\bibinfo{volume}{45}}, \bibinfo{pages}{75--79}
	(\bibinfo{year}{1980}).
	
	\bibitem{RN272}
	\bibinfo{author}{Caves, C.~M.}
	\newblock \bibinfo{title}{Quantum-mechanical noise in an interferometer}.
	\newblock \emph{\bibinfo{journal}{Phys. Rev. D}} \textbf{\bibinfo{volume}{23}},
	\bibinfo{pages}{1693--1708} (\bibinfo{year}{1981}).
	
	\bibitem{RN447}
	\bibinfo{author}{Yurke, B.}, \bibinfo{author}{McCall, S.~L.} \&
	\bibinfo{author}{Klauder, J.~R.}
	\newblock \bibinfo{title}{SU(2) and SU(1,1) interferometers}.
	\newblock \emph{\bibinfo{journal}{Phys. Rev. A}} \textbf{\bibinfo{volume}{33}},
	\bibinfo{pages}{4033--4054} (\bibinfo{year}{1986}).
	
	\bibitem{RN295}
	\bibinfo{author}{Giovannetti, V.}, \bibinfo{author}{Lloyd, S.} \&
	\bibinfo{author}{Maccone, L.}
	\newblock \bibinfo{title}{Quantum metrology}.
	\newblock \emph{\bibinfo{journal}{Phys Rev Lett}}
	\textbf{\bibinfo{volume}{96}}, \bibinfo{pages}{010401}
	(\bibinfo{year}{2006}).
	
	\bibitem{RN279}
	\bibinfo{author}{Holland, M.~J.} \& \bibinfo{author}{Burnett, K.}
	\newblock \bibinfo{title}{Interferometric detection of optical phase shifts at
		the Heisenberg limit}.
	\newblock \emph{\bibinfo{journal}{Phys. Rev. Lett.}}
	\textbf{\bibinfo{volume}{71}}, \bibinfo{pages}{1355--1358}
	(\bibinfo{year}{1993}).
	
	\bibitem{RN293}
	\bibinfo{author}{Beltr\'an, J.} \& \bibinfo{author}{Luis, A.}
	\newblock \bibinfo{title}{Breaking the Heisenberg limit with inefficient
		detectors}.
	\newblock \emph{\bibinfo{journal}{Phys. Rev. A}} \textbf{\bibinfo{volume}{72}}
	(\bibinfo{year}{2005}).
	
	\bibitem{RN292}
	\bibinfo{author}{Roy, S.~M.} \& \bibinfo{author}{Braunstein, S.~L.}
	\newblock \bibinfo{title}{Exponentially enhanced quantum metrology}.
	\newblock \emph{\bibinfo{journal}{Phys. Rev. Lett.}}
	\textbf{\bibinfo{volume}{100}} (\bibinfo{year}{2008}).
	
	\bibitem{RN277}
	\bibinfo{author}{Zwierz, M.}, \bibinfo{author}{Perez-Delgado, C.~A.} \&
	\bibinfo{author}{Kok, P.}
	\newblock \bibinfo{title}{General optimality of the Heisenberg limit for
		quantum metrology}.
	\newblock \emph{\bibinfo{journal}{Phys. Rev. Lett.}}
	\textbf{\bibinfo{volume}{105}}, \bibinfo{pages}{180402}
	(\bibinfo{year}{2010}).
	
	\bibitem{RN280}
	\bibinfo{author}{Hou, Z.} \emph{et~al.}
	\newblock \bibinfo{title}{``super-Heisenberg" and Heisenberg scalings achieved
		simultaneously in the estimation of a rotating field}.
	\newblock \emph{\bibinfo{journal}{Phys. Rev. Lett.}}
	\textbf{\bibinfo{volume}{126}}, \bibinfo{pages}{070503}
	(\bibinfo{year}{2021}).
	
	\bibitem{RN469}
	\bibinfo{author}{Yang, J.}, \bibinfo{author}{Pang, S.}, \bibinfo{author}{del
		Campo, A.} \& \bibinfo{author}{Jordan, A.~N.}
	\newblock \bibinfo{title}{Super-Heisenberg scaling in Hamiltonian parameter
		estimation in the long-range Kitaev chain}.
	\newblock \emph{\bibinfo{journal}{Phys. Rev. Res.}}
	\textbf{\bibinfo{volume}{4}} (\bibinfo{year}{2022}).
	
	\bibitem{RN644}
	\bibinfo{author}{Erker, P.} \emph{et~al.}
	\newblock \bibinfo{title}{Autonomous quantum clocks: Does thermodynamics limit
		our ability to measure time?}
	\newblock \emph{\bibinfo{journal}{Phys. Rev. X}} \textbf{\bibinfo{volume}{7}}
	(\bibinfo{year}{2017}).
	
	\bibitem{RN643}
	\bibinfo{author}{Pearson, A.~N.} \emph{et~al.}
	\newblock \bibinfo{title}{Measuring the thermodynamic cost of timekeeping}.
	\newblock \emph{\bibinfo{journal}{Phys. Rev. X}} \textbf{\bibinfo{volume}{11}}
	(\bibinfo{year}{2021}).
	
	\bibitem{RN648}
	\bibinfo{author}{Allan, D.~W.}
	\newblock \bibinfo{title}{Statistics of atomic frequency standards}.
	\newblock \emph{\bibinfo{journal}{Proc. IEEE}} \textbf{\bibinfo{volume}{54}},
	\bibinfo{pages}{221--230} (\bibinfo{year}{1966}).
	
	\bibitem{RN153}
	\bibinfo{author}{Bollinger, J.~J.}, \bibinfo{author}{Itano, W.~M.},
	\bibinfo{author}{Wineland, D.~J.} \& \bibinfo{author}{Heinzen, D.~J.}
	\newblock \bibinfo{title}{Optimal frequency measurements with maximally
		correlated states}.
	\newblock \emph{\bibinfo{journal}{Phys. Rev. A}} \textbf{\bibinfo{volume}{54}},
	\bibinfo{pages}{R4649--R4652} (\bibinfo{year}{1996}).
	
	\bibitem{RN249}
	\bibinfo{author}{Ockeloen, C.~F.}, \bibinfo{author}{Schmied, R.},
	\bibinfo{author}{Riedel, M.~F.} \& \bibinfo{author}{Treutlein, P.}
	\newblock \bibinfo{title}{Quantum metrology with a scanning probe atom
		interferometer}.
	\newblock \emph{\bibinfo{journal}{Phys. Rev. Lett.}}
	\textbf{\bibinfo{volume}{111}} (\bibinfo{year}{2013}).
	
	\bibitem{RN248}
	\bibinfo{author}{Muessel, W.}, \bibinfo{author}{Strobel, H.},
	\bibinfo{author}{Linnemann, D.}, \bibinfo{author}{Hume, D.~B.} \&
	\bibinfo{author}{Oberthaler, M.~K.}
	\newblock \bibinfo{title}{Scalable spin squeezing for quantum-enhanced
		magnetometry with Bose-Einstein condensates}.
	\newblock \emph{\bibinfo{journal}{Phys. Rev. Lett.}}
	\textbf{\bibinfo{volume}{113}} (\bibinfo{year}{2014}).
	
	\bibitem{RN267}
	\bibinfo{author}{Hosten, O.}, \bibinfo{author}{Krishnakumar, R.},
	\bibinfo{author}{Engelsen, N.~J.} \& \bibinfo{author}{Kasevich, M.~A.}
	\newblock \bibinfo{title}{Quantum phase magnification}.
	\newblock \emph{\bibinfo{journal}{Science}} \textbf{\bibinfo{volume}{352}},
	\bibinfo{pages}{1552--1555} (\bibinfo{year}{2016}).
	
	\bibitem{RN71}
	\bibinfo{author}{Degen, C.~L.}, \bibinfo{author}{Reinhard, F.} \&
	\bibinfo{author}{Cappellaro, P.}
	\newblock \bibinfo{title}{Quantum sensing}.
	\newblock \emph{\bibinfo{journal}{Rev. Mod. Phys.}}
	\textbf{\bibinfo{volume}{89}} (\bibinfo{year}{2017}).
	
	\bibitem{RN301}
	\bibinfo{author}{Wootters, W.~K.}
	\newblock \bibinfo{title}{Statistical distance and Hilbert-space}.
	\newblock \emph{\bibinfo{journal}{Phys. Rev. D}} \textbf{\bibinfo{volume}{23}},
	\bibinfo{pages}{357--362} (\bibinfo{year}{1981}).
	
	\bibitem{RN327}
	\bibinfo{author}{Anandan, J.} \& \bibinfo{author}{Aharonov, Y.}
	\newblock \bibinfo{title}{Geometry of quantum evolution}.
	\newblock \emph{\bibinfo{journal}{Phys. Rev. Lett.}}
	\textbf{\bibinfo{volume}{65}}, \bibinfo{pages}{1697--1700}
	(\bibinfo{year}{1990}).
	
	\bibitem{RN288}
	\bibinfo{author}{Braunstein, S.~L.}, \bibinfo{author}{Caves, C.~M.} \&
	\bibinfo{author}{Milburn, G.~J.}
	\newblock \bibinfo{title}{Generalized uncertainty relations: Theory, examples,
		and lorentz invariance}.
	\newblock \emph{\bibinfo{journal}{Ann. Phys.}} \textbf{\bibinfo{volume}{247}},
	\bibinfo{pages}{135--173} (\bibinfo{year}{1996}).
	
	\bibitem{RN807}
	\bibinfo{author}{Meier, F.}, \bibinfo{author}{Schwarzhans, E.},
	\bibinfo{author}{Erker, P.} \& \bibinfo{author}{Huber, M.}
	\newblock \bibinfo{title}{Fundamental accuracy-resolution trade-off for
		timekeeping devices}.
	\newblock \emph{\bibinfo{journal}{Phys. Rev. Lett.}}
	\textbf{\bibinfo{volume}{131}}, \bibinfo{pages}{220201}
	(\bibinfo{year}{2023}).
	
	\bibitem{RN304}
	\bibinfo{author}{Giovannetti, V.}, \bibinfo{author}{Lloyd, S.} \&
	\bibinfo{author}{Maccone, L.}
	\newblock \bibinfo{title}{Quantum limits to dynamical evolution}.
	\newblock \emph{\bibinfo{journal}{Phys. Rev. A}} \textbf{\bibinfo{volume}{67}}
	(\bibinfo{year}{2003}).
	
	\bibitem{RN316}
	\bibinfo{author}{Giovannetti, V.}, \bibinfo{author}{Lloyd, S.} \&
	\bibinfo{author}{Maccone, L.}
	\newblock \bibinfo{title}{The role of entanglement in dynamical evolution}.
	\newblock \emph{\bibinfo{journal}{Europhys. Lett.}}
	\textbf{\bibinfo{volume}{62}}, \bibinfo{pages}{615--621}
	(\bibinfo{year}{2003}).
	
	\bibitem{RN370}
	\bibinfo{author}{Fr\"owis, F.}
	\newblock \bibinfo{title}{Kind of entanglement that speeds up quantum
		evolution}.
	\newblock \emph{\bibinfo{journal}{Phys. Rev. A}} \textbf{\bibinfo{volume}{85}}
	(\bibinfo{year}{2012}).
	
	\bibitem{RN291}
	\bibinfo{author}{Jones, P.~J.} \& \bibinfo{author}{Kok, P.}
	\newblock \bibinfo{title}{Geometric derivation of the quantum speed limit}.
	\newblock \emph{\bibinfo{journal}{Phys. Rev. A}} \textbf{\bibinfo{volume}{82}}
	(\bibinfo{year}{2010}).
	
	\bibitem{RN361}
	\bibinfo{author}{Taddei, M.~M.}, \bibinfo{author}{Escher, B.~M.},
	\bibinfo{author}{Davidovich, L.} \& \bibinfo{author}{de~Matos, R.~L.}
	\newblock \bibinfo{title}{Quantum speed limit for physical processes}.
	\newblock \emph{\bibinfo{journal}{Phys. Rev. Lett.}}
	\textbf{\bibinfo{volume}{110}} (\bibinfo{year}{2013}).
	
	\bibitem{RN401}
	\bibinfo{author}{Pires, D.~P.}, \bibinfo{author}{Cianciaruso, M.},
	\bibinfo{author}{Celeri, L.~C.}, \bibinfo{author}{Adesso, G.} \&
	\bibinfo{author}{Soares-Pinto, D.~O.}
	\newblock \bibinfo{title}{Generalized geometric quantum speed limits}.
	\newblock \emph{\bibinfo{journal}{Phys. Rev. X}} \textbf{\bibinfo{volume}{6}}
	(\bibinfo{year}{2016}).
	
	\bibitem{RN402}
	\bibinfo{author}{Gessner, M.} \& \bibinfo{author}{Smerzi, A.}
	\newblock \bibinfo{title}{Statistical speed of quantum states: Generalized
		quantum fisher information and Schatten speed}.
	\newblock \emph{\bibinfo{journal}{Phys. Rev. A}} \textbf{\bibinfo{volume}{97}}
	(\bibinfo{year}{2018}).
	
	\bibitem{RN390}
	\bibinfo{author}{Garc\'ia-Pintos, L.~P.}, \bibinfo{author}{Nicholson, S.~B.},
	\bibinfo{author}{Green, J.~R.}, \bibinfo{author}{del Campo, A.} \&
	\bibinfo{author}{Gorshkov, A.~V.}
	\newblock \bibinfo{title}{Unifying quantum and classical speed limits on
		observables}.
	\newblock \emph{\bibinfo{journal}{Phys. Rev. X}} \textbf{\bibinfo{volume}{12}}
	(\bibinfo{year}{2022}).
	
	\bibitem{RN537}
	\bibinfo{author}{Huelga, S.~F.} \emph{et~al.}
	\newblock \bibinfo{title}{Improvement of frequency standards with quantum
		entanglement}.
	\newblock \emph{\bibinfo{journal}{Phys. Rev. Lett.}}
	\textbf{\bibinfo{volume}{79}}, \bibinfo{pages}{3865--3868}
	(\bibinfo{year}{1997}).
	
	\bibitem{RN951}
	\bibinfo{author}{Scully, M.~O.} \& \bibinfo{author}{Zubairy, M.~S.}
	\newblock \emph{\bibinfo{title}{Quantum Optics}} (\bibinfo{publisher}{Cambridge
		University Press}, \bibinfo{address}{Cambridge}, \bibinfo{year}{1997}).
	
	\bibitem{RN949}
	\bibinfo{author}{Hudelist, F.} \emph{et~al.}
	\newblock \bibinfo{title}{Quantum metrology with parametric amplifier-based
		photon correlation interferometers}.
	\newblock \emph{\bibinfo{journal}{Nat. Commun.}} \textbf{\bibinfo{volume}{5}},
	\bibinfo{pages}{3049} (\bibinfo{year}{2014}).
	
	\bibitem{RN950}
	\bibinfo{author}{Manceau, M.}, \bibinfo{author}{Leuchs, G.},
	\bibinfo{author}{Khalili, F.} \& \bibinfo{author}{Chekhova, M.}
	\newblock \bibinfo{title}{Detection loss tolerant supersensitive phase
		measurement with an SU(1,1) interferometer}.
	\newblock \emph{\bibinfo{journal}{Phys. Rev. Lett.}}
	\textbf{\bibinfo{volume}{119}}, \bibinfo{pages}{223604}
	(\bibinfo{year}{2017}).
	
	\bibitem{RN726}
	\bibinfo{author}{Liu, Q.} \emph{et~al.}
	\newblock \bibinfo{title}{Nonlinear interferometry beyond classical limit
		enabled by cyclic dynamics}.
	\newblock \emph{\bibinfo{journal}{Nat. Phys.}} \textbf{\bibinfo{volume}{18}},
	\bibinfo{pages}{167–171} (\bibinfo{year}{2022}).
	
	\bibitem{RN947}
	\bibinfo{author}{Linnemann, D.} \emph{et~al.}
	\newblock \bibinfo{title}{Quantum-enhanced sensing based on time reversal of
		nonlinear dynamics}.
	\newblock \emph{\bibinfo{journal}{Phys. Rev. Lett.}}
	\textbf{\bibinfo{volume}{117}}, \bibinfo{pages}{013001}
	(\bibinfo{year}{2016}).
	
	\bibitem{RN532}
	\bibinfo{author}{Napolitano, M.} \emph{et~al.}
	\newblock \bibinfo{title}{Interaction-based quantum metrology showing scaling
		beyond the Heisenberg limit}.
	\newblock \emph{\bibinfo{journal}{Nature}} \textbf{\bibinfo{volume}{471}},
	\bibinfo{pages}{486--9} (\bibinfo{year}{2011}).
	
	\bibitem{RN826}
	\bibinfo{author}{Napolitano, M.} \& \bibinfo{author}{Mitchell, M.~W.}
	\newblock \bibinfo{title}{Nonlinear metrology with a quantum interface}.
	\newblock \emph{\bibinfo{journal}{New J. Phys.}} \textbf{\bibinfo{volume}{12}}
	(\bibinfo{year}{2010}).
	
	\bibitem{RN948}
	\bibinfo{author}{Guo, Y.-F.}, \bibinfo{author}{Zhong, W.},
	\bibinfo{author}{Zhou, L.} \& \bibinfo{author}{Sheng, Y.-B.}
	\newblock \bibinfo{title}{Supersensitivity of Kerr phase estimation with
		two-mode squeezed vacuum states}.
	\newblock \emph{\bibinfo{journal}{Phys. Rev. A}}
	\textbf{\bibinfo{volume}{105}}, \bibinfo{pages}{032609}
	(\bibinfo{year}{2022}).
	
	\bibitem{RN1057}
	\bibinfo{author}{Rubio, J.} \& \bibinfo{author}{Dunningham, J.}
	\newblock \bibinfo{title}{Bayesian multiparameter quantum metrology with
		limited data}.
	\newblock \emph{\bibinfo{journal}{Phys. Rev. A}}
	\textbf{\bibinfo{volume}{101}}, \bibinfo{pages}{032114}
	(\bibinfo{year}{2020}).
	
	\bibitem{RN1058}
	\bibinfo{author}{Rubio, J.} \& \bibinfo{author}{Dunningham, J.}
	\newblock \bibinfo{title}{Quantum metrology in the presence of limited data}.
	\newblock \emph{\bibinfo{journal}{New J. Phys.}} \textbf{\bibinfo{volume}{21}},
	\bibinfo{pages}{043037} (\bibinfo{year}{2019}).
	
	\bibitem{RN604}
	\bibinfo{author}{Jarzyna, M.} \& \bibinfo{author}{Demkowicz-Dobrzański, R.}
	\newblock \bibinfo{title}{True precision limits in quantum metrology}.
	\newblock \emph{\bibinfo{journal}{New J. Phys.}} \textbf{\bibinfo{volume}{17}}
	(\bibinfo{year}{2015}).
	
	\bibitem{RN281}
	\bibinfo{author}{Gorecki, W.}, \bibinfo{author}{Demkowicz-Dobrzanski, R.},
	\bibinfo{author}{Wiseman, H.~M.} \& \bibinfo{author}{Berry, D.~W.}
	\newblock \bibinfo{title}{pi-corrected Heisenberg limit}.
	\newblock \emph{\bibinfo{journal}{Phys. Rev. Lett.}}
	\textbf{\bibinfo{volume}{124}} (\bibinfo{year}{2020}).
	
	\bibitem{RN889}
	\bibinfo{author}{Berry, D.~W.} \& \bibinfo{author}{Wiseman, H.~M.}
	\newblock \bibinfo{title}{Optimal states and almost optimal adaptive
		measurements for quantum interferometry}.
	\newblock \emph{\bibinfo{journal}{Phys. Rev. Lett.}}
	\textbf{\bibinfo{volume}{85}}, \bibinfo{pages}{5098--5101}
	(\bibinfo{year}{2000}).
	
	\bibitem{RN922}
	\bibinfo{author}{Giovannetti, V.} \& \bibinfo{author}{Maccone, L.}
	\newblock \bibinfo{title}{Sub-Heisenberg estimation strategies are
		ineffective}.
	\newblock \emph{\bibinfo{journal}{Phys. Rev. Lett.}}
	\textbf{\bibinfo{volume}{108}}, \bibinfo{pages}{210404}
	(\bibinfo{year}{2012}).
	
	\bibitem{RN303}
	\bibinfo{author}{Tsang, M.}
	\newblock \bibinfo{title}{Ziv-Zakai error bounds for quantum parameter
		estimation}.
	\newblock \emph{\bibinfo{journal}{Phys. Rev. Lett.}}
	\textbf{\bibinfo{volume}{108}}, \bibinfo{pages}{230401}
	(\bibinfo{year}{2012}).
	
	\bibitem{RN252}
	\bibinfo{author}{Liu, J.}, \bibinfo{author}{Yuan, H.~D.}, \bibinfo{author}{Lu,
		X.~M.} \& \bibinfo{author}{Wang, X.~G.}
	\newblock \bibinfo{title}{Quantum Fisher information matrix and multiparameter
		estimation}.
	\newblock \emph{\bibinfo{journal}{J. Phys. A: Math. Theor.}}
	\textbf{\bibinfo{volume}{53}} (\bibinfo{year}{2020}).
	
\end{thebibliography}
\end{document}